# Social signals dominate clinician trust in medical AI

**Aníbal Astobiza**[1,*], **Ramón Ortega Lozano**[2], **Marcos Alonso**[3], **José Ríos-Díaz**[2]
[1]Department of Philosophy I, University of Granada
[2]San Juan de Dios Foundation · SJD School of Nursing and Physiotherapy · Comillas Pontifical University, Madrid, Spain
[3]Complutense University of Madrid
amastobiza@ugr.es; rortegal@comillas.edu; marcos.alonso@ucm.es; jriosd@comillas.edu
* *Corresponding author*

**Abstract**

**Objective:** Dependence on automation should primarily be influenced by accuracy and outcome data, as prevailing theories of trust suggest that trust in automation is largely based on demonstrated performance. We test this assumption against an alternative account on which clinicians evaluate artificial intelligence (AI) with the same social-cognitive machinery they use for human agents and treat trustworthiness as a latent property inferred from how a system behaves and is embedded in practice as well as from what it achieves. We introduce SPOT (Synergistic Process-Outcome Trust), a Bayesian framework that models trust as an inference over latent agent properties such as competence and benevolence, drawn from both outcome and process signals.

**Method:** In a within-subjects vignette study, 275 Spanish healthcare professionals evaluated 16 clinical AI scenarios that systematically varied outcome signals (accuracy, intervention timing, prognosis) and process signals (interface familiarity, explainability, team consensus, supervision, personalization). We analysed the ratings with linear mixed models complemented by participant-wise cross-validation.

**Results:** Process signals dominated trust formation. Interface familiarity (β = 0.71), team consensus (β = 0.64), and explainability (β = 0.63) produced large effects ($p < .001$). Stated accuracy produced a statistically reliable yet practically small effect (β = 0.10). Aggregating absolute effect magnitudes, process signals accounted for 80.3% of the total and outcome signals for 19.7%, an approximately four-fold inversion of the rational-agent prediction. Trust baselines were highly idiosyncratic and poorly recovered from demographics ($R^2$ = 0.06 under group-wise cross-validation).

**Conclusions:** Clinicians adopt an intentional-stance heuristic and treat AI systems as agents whose trustworthiness needs social validation as well as technical verification. Building trustworthy clinical AI is therefore a social and institutional task as well as a technical one.

## 1. Introduction

The integration of artificial intelligence (AI) into clinical practice marks one of the most consequential shifts in modern medicine. AI systems now interpret diagnostic images, help triage patients, and recommend treatments, and they promise to augment and in some workflows partially automate clinical decision-making. A technically sound deployment still has to be used before it can help anyone. The decisive question is behavioural. Under what conditions will healthcare professionals trust these systems enough to incorporate their recommendations into the care of real patients? This question sits at the intersection of medical informatics and cognitive science, because it concerns how humans form judgments about the reliability and intentions of artificial agents under uncertainty.

Trust in artificial agents has most often been modelled as a function of observed performance. According to this perspective, humans function as decision-makers who are roughly calibrated, updating their trust primarily based on outcome feedback. They tend to increase their reliance on a system when it performs well and decrease it when the system makes mistakes (Hidalgo et al., 2021). Such outcome-dominant accounts presuppose conditions that laboratory automation often meets and that clinical AI routinely violates. These conditions are repeated interaction, relatively low stakes, and feedback that is timely and easy to interpret. In real clinical deployments, ground truth is often delayed, confounded, or never observed; the stakes are high; and a single salient error can be catastrophic. Under sparse outcome information and high epistemic risk, decision-makers rely on cues beyond a track record to judge whether a system is competent and well-intentioned. We therefore treat trust in clinical AI as social inference under uncertainty, and we introduce SPOT (Synergistic Process-Outcome Trust), a Bayesian framework that integrates outcome evidence with social and procedural signals such as consensus, transparency, and explainability.

Trust is a multidimensional construct. It involves a willingness to accept vulnerability based on positive expectations about another party's future behaviour (Rousseau et al., 1998). Influential models in organizational psychology decompose perceived trustworthiness into distinct antecedents, mainly ability or competence, benevolence, and integrity, which observers infer from limited evidence (Mayer et al., 1995). These antecedents are latent. Observers never see them directly and have to infer them from cues. The human-factors literature on trust in automation extends this logic to non-human agents and frames the engineering goal as appropriate reliance, meaning trust calibrated to a system's true capabilities so that users avoid both over-reliance on fallible automation and under-use of reliable automation (Lee & See, 2004). In order to assess trust in this manner, it is necessary to have diagnostic insights into the underlying skills or competences involved. Sadly, this crucial knowledge is frequently lacking when crucial therapeutic judgements need to be made.

Dennett's (1987) concept of the intentional stance offers a parsimonious account of how humans cope with this scarcity. The intentional stance is the strategy of treating a complex system as if it were a rational agent with beliefs and desires, so as to predict its behaviour without modelling its underlying mechanism. The stance is computationally efficient, and people adopt it readily toward artefacts. A clinician who adopts it toward AI may evaluate the system with the same cognitive machinery she applies to human colleagues, including the heuristics, social cues, and norms that govern interpersonal trust. That machinery differs from the dispassionate cost-benefit calculus that rational-agent models prescribe.

Computational cognitive science gives this intuition a formal backbone. A large body of work characterizes many facets of human reasoning as approximate Bayesian inference over structured probabilistic models of the world (Tenenbaum et al., 2011; Griffiths et al., 2010). Within this tradition, intuitive psychology has been formalized as inverse planning. Observers model other agents as approximately rational planners who choose actions to maximize expected utility, and they then invert this generative model to infer the mental states that best explain the observed behaviour, including beliefs, desires, and competence (Baker et al., 2009; Jara-Ettinger et al., 2016). The same machinery that lets us infer a colleague's intentions from their actions can in principle be turned on an algorithm. A clinician who sees that an AI is endorsed by trusted peers, explains its reasoning, and sits within legitimate governance is using socially and procedurally diagnostic cues to infer a latent property, trustworthiness, under genuine uncertainty. This behaviour fits broader computational-rationality perspectives, in which intelligent agents allocate limited inferential resources to reach good-enough decisions given the costs of computation (Gershman et al., 2015).

Empirical research on human-AI interaction has documented two tendencies that any adequate account must reconcile. Users frequently display automation bias. They over-rely on automated recommendations and miss errors that they would have caught unaided (Goddard et al., 2012). Users also display algorithm aversion. They prefer human judgment and abandon algorithms after seeing them err, even when the algorithms outperform humans (Dietvorst et al., 2015). Under other framings people show the opposite pattern, called algorithm appreciation, and weight algorithmic advice more heavily than identical human advice (Logg et al., 2019). In medical settings specifically, patients and professionals can resist AI because they perceive their own case as unique and therefore poorly served by statistical reasoning (Longoni et al., 2019).

The coexistence of aversion and appreciation indicates that reliance shifts with contextual cues about the agent and the decision. In healthcare, acceptance has been associated with transparency, perceived usefulness, workflow fit, and organizational support (Asan et al., 2020; Greenhalgh et al., 2017). Much of this evidence is descriptive and correlational, which makes it hard to isolate the causal contribution of any single cue. A controlled factorial manipulation of outcome and process signals is needed to adjudicate between outcome-dominant and process-sensitive accounts.

### *1.1. The SPOT framework*

SPOT distinguishes two broad categories of trust-relevant signal. Outcome signals (O) reflect demonstrated or stated performance, such as diagnostic accuracy, expected clinical benefit, and prognostic content. Classical rational-agent models hold that these signals should dominate trust formation, because a consistent decision-maker ought to weight a recommendation chiefly by its probability of producing good outcomes. Process signals (P) reflect how the system operates and how it is embedded within clinical practice. We organize them into three conceptual subcategories. Cognitive transparency concerns whether the clinician can understand the system’s reasoning. Emotional comfort concerns whether using the system feels appropriate and familiar. Institutional legitimacy concerns whether trusted colleagues endorse the system and recognized authorities validate it.

We formalize trust as a weighted combination of the two signal classes, $T \approx \alpha \cdot f(P) + \beta \cdot g(O)$, where $\alpha$ and $\beta$ are structural weights on processed process and outcome information. Rational-agent models motivate the hypothesis of outcome dominance ($\beta > \alpha$). Process signals could nonetheless carry substantial weight for at least three reasons. First, they may serve as proxies for outcome quality when direct performance information is unavailable. Second, they may track legitimate values beyond consequences, such as understanding, autonomy, and professional integrity. Third, clinicians may reason about AI as an intentional agent whose trustworthiness depends on its character as much as on its track record.

These possibilities can be expressed within a single generative scheme. Let Z denote latent trustworthiness. Trust formation is then Bayesian inference over Z given outcome cues O and process cues P, with $P(Z \mid O, P) \propto P(O \mid Z) \cdot P(P \mid Z) \cdot P(Z)$. Here the prior $P(Z)$ is shaped by individual differences such as AI familiarity and professional experience, while $P(O \mid Z)$ and $P(P \mid Z)$ are the likelihoods of observing each cue given a system's true trustworthiness. The relative influence of a cue on the posterior depends on its diagnosticity, that is, on how sharply it discriminates trustworthy from untrustworthy systems. Suppose clinicians' intuitive theories hold that only trustworthy systems are transparent and institutionally endorsed, which gives a sharp $P(P \mid Z)$. Process cues then move the posterior more than a nominal accuracy figure whose likelihood is comparatively flat or hard to interpret. The present study does not estimate this hierarchical Bayesian model directly. We instead use linear mixed models to recover the empirical effect magnitude of each manipulated cue, and we treat the relative magnitude of process and outcome coefficients as a behavioural index of how clinicians weight these signals. A full hierarchical implementation is left to future work.

## 2. Method

### *2.1. Participants*

The analytic sample comprised 275 Spanish healthcare professionals. Inclusion required completion of at least one experimental vignette and self-reported sex coded as male or female. Two cases with non-binary coding were excluded from the comparative analyses, and an alternative analysis that retains them is reported in the Supplementary Materials and yields equivalent conclusions. Of the 326 individuals who began the survey, 48 provided no vignette responses and were treated as non-completers, and a further 3 completers lacked sex information. Excluding these cases yielded the final analytic sample of N = 275. Of these, 215 (78.2%) provided complete data for all 16 vignettes, and the remainder gave partial responses (mean = 13.6 vignettes per participant). The mixed-model analyses use all 3,744 available trust ratings under a missing-at-random assumption, because maximum-likelihood estimation for mixed models accommodates unbalanced data without listwise deletion.

The modal age stratum was 46–55 years. Women constituted 58.9% (n = 162) of the sample and men 41.1% (n = 113). Physicians were the most represented profession (68.0%), followed by physiotherapists (12.7%), nurses (7.6%), psychologists and clinical psychologists (6.9%), pharmacists (4.0%), and dentists (0.7%). Professional experience was concentrated in senior strata, with 49.5% reporting more than 20 years and 21.8% reporting 11–20 years. Self-reported familiarity with AI was modest, with 39.6% "moderately familiar," 38.2% "slightly familiar," 12.0% "not at all familiar," 8.4% "very familiar," and 1.8% "extremely familiar." Anonymized data and analysis scripts are openly available at the Open Science Framework (OSF).

### *2.2. Materials and design*

We used a within-subjects vignette design (Aguinis & Bradley, 2014). Each participant evaluated 16 clinical scenarios organized into 8 matched pairs, with each pair isolating a single factor while holding the others constant. The scenarios spanned diagnostic decision-making, treatment planning, monitoring, and prognostic assessment, and they were developed in consultation with practising physicians to maximize clinical realism.

The eight experimental factors comprised three outcome signals and five process signals. The outcome signals were *accuracy*, set to high (95%) or moderate (80%); *intervention timing*, immediate or progressive; and *prognosis*, favourable or adverse. The process signals were *interface*, traditional or innovative; *explainability*, a detailed explanation or a black box that no one could justify, including the AI itself, the physicians, and the engineers; *team consensus*, a team described as convinced or skeptical; *supervision and autonomy*, human-supervised or autonomous robotic operation; and *personalization*, a personalized recommendation or a standard protocol. Trust was rated on a 5-point Likert scale, from 1 (not at all trustworthy) to 5 (extremely trustworthy) (Jian et al., 2000).

### *2.3. Procedure and analysis*

The study was administered online via PsyToolkit (Stoet, 2010, 2017). After providing informed consent, participants completed a brief demographic questionnaire (profession, experience, age, sex, and AI familiarity) and then evaluated the vignettes in a fixed order. Paired vignettes were separated by at least three intervening items to reduce direct side-by-side comparison. Median completion time was approximately 6 minutes. The protocol received ethics approval from the University of Granada (4948/CEIH/2025).

We analysed trust ratings with linear mixed models (LMMs; statsmodels, Seabold & Perktold, 2010), specifying a by-participant random intercept to accommodate the repeated-measures structure. Fixed-effect coefficients (β) index the estimated change in trust attributable to each factor, holding participant baselines constant. Because the design is within-subjects and the contrasts are paired, each coefficient is closely approximated by the corresponding within-condition mean difference. To assess how well trust predictions generalize to new users, we additionally trained gradient-boosted and baseline regression models (including LightGBM; Ke et al., 2017) using Group K-Fold cross-validation (k = 5), ensuring that no participant appeared in both training and test folds. This procedure prevents the models from simply memorizing individual baselines and yields an estimate of true out-of-sample prediction.

### *2.4. Data provenance and secondary analysis*

The analyses reported here form a secondary analysis of a dataset that the authors collected for a broader programme on healthcare professionals' attitudes toward clinical AI. A separate study based on the same survey is in preparation, and it addresses different research questions with a different analytic strategy. We state this provenance openly so that readers can place the present work within that programme and avoid treating the two outputs as independent samples. The de-identified data, the survey instrument, and the analysis scripts are deposited at the Open Science Framework, and the companion work will cite the same deposit.

This article makes a distinct contribution. The primary research questions, the SPOT framework, the linear mixed-model estimation of cue weights, and the machine-learning test of baseline idiosyncrasy are specific to this paper and did not form part of the companion analysis. The text, the theoretical framing, and the discussion were written for this article alone. We report new estimands, namely the relative structural weights of process and outcome signals. We do not re-describe previously reported results with a different software package.

Reusing a dataset to answer new questions with new methods is an established and encouraged practice under Responsible Research and Innovation and Open Science initiaitves. The European Code of Conduct for Research Integrity treats data and metadata as legitimate, citable research products and asks that research data be as open as possible and as closed as necessary (ALLEA, 2023). The same principle underlies the FAIR guidelines for findable, accessible, interoperable, and reusable data (Wilkinson et al., 2016) and the UNESCO Recommendation on Open Science, which 193 member states adopted to promote efficient reuse of publicly funded research (UNESCO, 2021). National and institutional codes converge on the same standard. The CSIC code of good scientific practice separates redundant publication, which it prohibits, from value-adding secondary analysis, which it supports when authors declare the origin of the data (CSIC, 2021). The Max Planck Society rules of good scientific practice ask that researchers exploit complex datasets fully and guard against fragmentary publication (Max Planck Society, 2021). Large research infrastructures institutionalize the practice, and the CERN open-data policy releases experimental data for continued analysis with new algorithms and models and requires only that users cite the original dataset (CERN, 2020). Methodological guidance for the health and behavioural sciences sets out when secondary analysis adds value, namely when it poses new questions, applies new techniques, or examines a defined subgroup in depth (Cheng & Phillips, 2014).

We fulfill the requirements and responsibilities dictated by these frameworks. We cite the data deposit, we will cite the companion study once it is available, we state the provenance of the data here, and we show throughout the Results and Discussion how the SPOT analysis reaches conclusions that the companion work does not address. We also took care to avoid duplicating text, figures, or conclusions across the two outputs.

## 3. Results

### *3.1. Distribution of trust ratings*

Across all vignettes and participants, trust ratings showed a clear central tendency (Figure 1). The total number of observations was N = 3,744. The modal response was "moderately trustworthy" (3), accounting for 45.0% of responses (n = 1,685). "Very trustworthy" (4) comprised 28.6% (n = 1,072) and "slightly trustworthy" (2) accounted for 17.0% (n = 636). The extremes were uncommon. "Not at all trustworthy" (1) represented 5.2% (n = 194) and "extremely trustworthy" (5) represented 4.2% (n = 157). Mean trust was M = 3.10 (SD = 0.91, 95% CI [3.07, 3.13]), and the distribution showed a slight negative skew (skewness = −0.27, kurtosis = −0.06), consistent with guarded optimism.

This pattern of moderate, guarded optimism converges with qualitative accounts of AI adoption in healthcare. Clinicians do not defer blindly to AI, and they do not reject it reflexively. They suspend judgment and stay sensitive to contextual cues. The concentration around the scale midpoint likely reflects genuine uncertainty about AI trustworthiness in ambiguous scenarios, reinforced by a professional ethos of prudence.

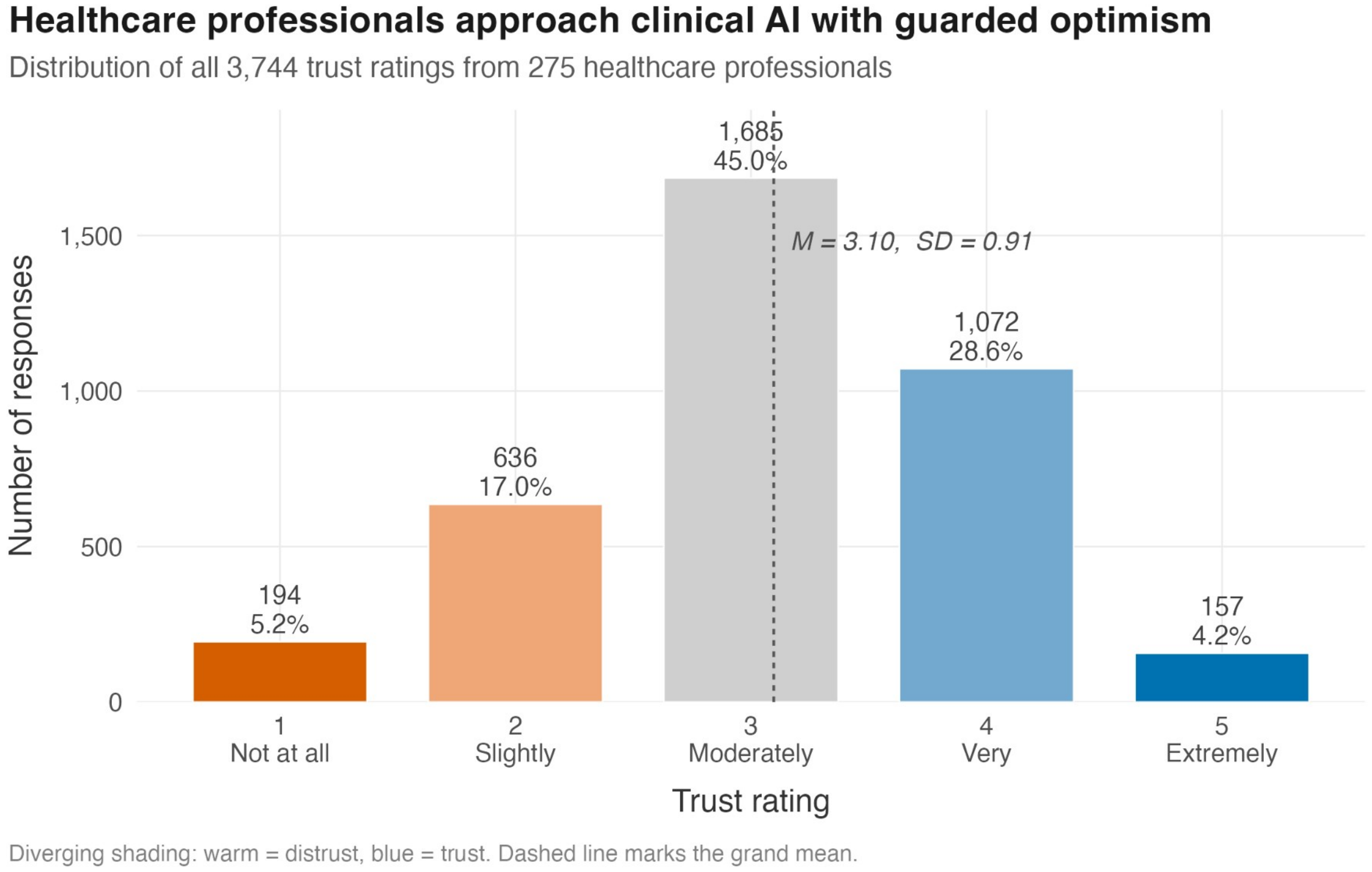

**Figure 1.** Overall distribution of trust ratings across all 16 vignettes and 275 participants (N = 3,744 observations). Bars show the frequency and percentage for each Likert response (1 = not at all trustworthy; 5 = extremely trustworthy), and the diverging shading runs from distrust (warm) to trust (blue). The distribution is centred at the mode of 3 (“moderately trustworthy”, 45.0%) with a slight negative skew (M = 3.10, SD = 0.91), which indicates a baseline of cautious trust.

### *3.2. Effects of the experimental factors*

The manipulation of process and outcome signals produced markedly different effects on trust (Table 1; Figure 2). Consistent with SPOT, process signals elicited the strongest responses. *Interface design* showed the largest effect (β = 0.71, 95% CI [0.58, 0.84], $p < .001$). Traditional, familiar interfaces drew substantially higher trust than innovative ones, holding participant baselines constant. Familiarity thus worked as a powerful heuristic, and it appeared to operate independently of underlying performance.

*Team consensus* produced the second-largest effect (β = 0.64, 95% CI [0.52, 0.77], $p < .001$). When the description presented the medical team as convinced of the system's value, trust rose sharply compared with the skeptical condition, which underscores the weight of social proof. *Explainability* showed a comparably large effect (β = 0.63, 95% CI [0.50, 0.76], $p < .001$). Systems that offered a detailed explanation drew far higher trust than black-box systems whose recommendations no one could justify. Figure 3 displays the full response distributions for these three dominant process signals.

Stated technical *accuracy* produced a statistically reliable yet practically small effect (β = 0.10, 95% CI [0.04, 0.17], $p = .003$). Reporting 95% accuracy instead of 80% did increase trust, though by only about one-seventh of the boost that interface familiarity conferred. This pattern challenges the assumption that communicating accuracy metrics is the principal lever of clinical adoption. *Prognosis* produced a robust effect in the direction that clinical calibration predicts (β = 0.38 for the contrast adverse versus favourable, 95% CI [0.25, 0.50], $p < .001$). Participants trusted the system less when it offered an optimistic prognosis for a condition known to carry poor outcomes, and they plausibly read the optimism as miscalibration. The remaining factors, namely autonomy, personalization, and intervention timing, produced small effects, and the last of these did not reach significance.

**Table 1.** Fixed effects of the experimental factors on trust ratings (linear mixed model).

| **Factor** | **Comparison** | ***β*** | **95% CI** | ***p*** |
|---|---|---|---|---|
| Interface | Traditional > Innovative | 0.71 | [0.58, 0.84] | < .001 |
| Consensus | Convinced > Skeptical | 0.64 | [0.52, 0.77] | < .001 |
| Explainability | Explainable > Black-box | 0.63 | [0.50, 0.76] | < .001 |
| Prognosis | Adverse > Favorable | 0.38 | [0.25, 0.50] | < .001 |
| Autonomy | Human supervision > Robotic | 0.12 | [0.02, 0.21] | .031 |
| Accuracy | High (95%) > Moderate (80%) | 0.10 | [0.04, 0.17] | .003 |
| Personalization | Personalized > Standard | 0.10 | [0.01, 0.19] | .033 |
| Intervention | Immediate > Progressive | 0.06 | [−0.01, 0.13] | .092 |

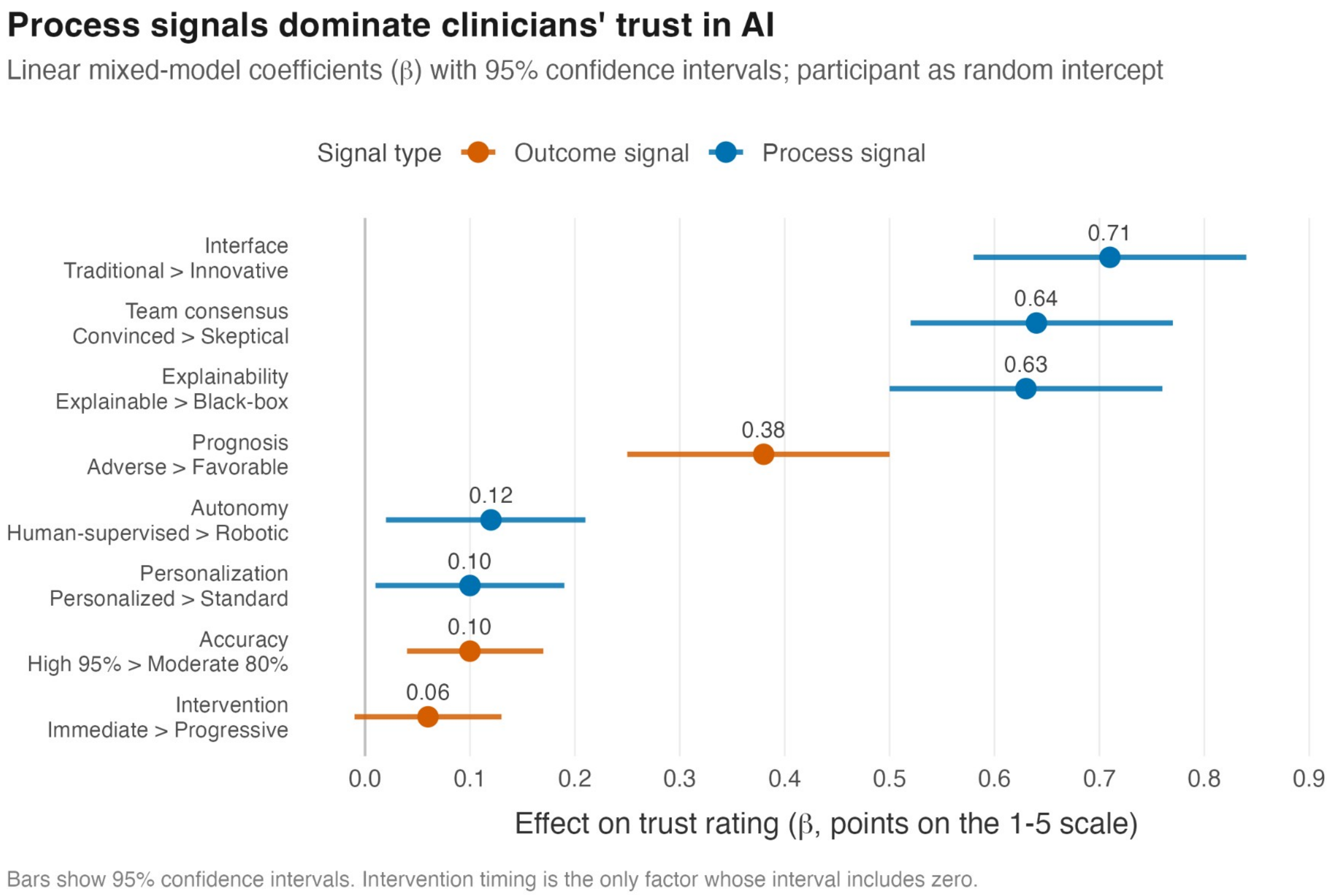

**Figure 2.** Fixed effects of the contextual factors on trust, estimated via linear mixed models. Points are the β coefficient (change in trust rating) relative to baseline, and bars are 95% confidence intervals. Process cues (blue) substantially outweigh outcome cues (vermillion), and intervention timing is the only factor whose interval includes zero.

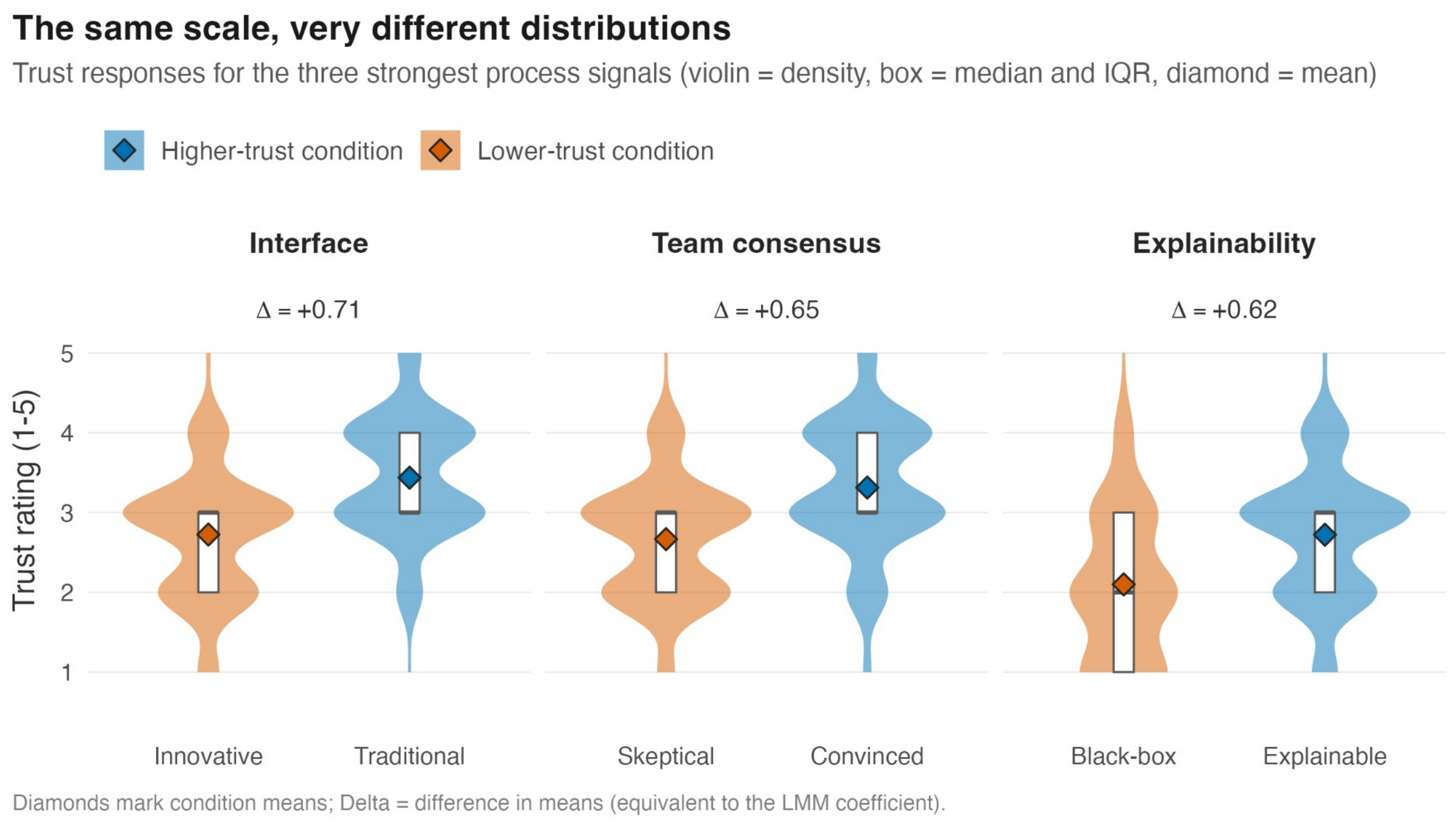

**Figure 3.** Distribution of trust ratings for the three dominant process signals. Within each panel, violins show the response density, the inner box marks the median and interquartile range, and the diamond marks the condition mean. Δ is the difference in condition means, which is equivalent to the LMM coefficient. Higher-trust conditions appear in blue and lower-trust conditions in vermillion.

### *3.3. Individual differences*

Beyond the experimental manipulations, participant characteristics were related to trust. AI familiarity correlated positively with trust at the observation level (Kendall's $\tau = 0.121$, $p < .001$) and more strongly at the participant level ($\tau = 0.221$, $p < .001$ over participant-mean trust). The relationship was monotonic. Participant-level mean trust rose from M = 2.91 (95% CI [2.72, 3.10], n = 33) among those "not at all familiar" to M = 3.59 (95% CI [3.27, 3.91], n = 5) among those "extremely familiar" (Figure 4). In the mixed model, familiarity predicted participant random intercepts and did not significantly moderate the slope of any experimental factor. This pattern suggests that familiarity shapes the prior P(Z) and raises baseline trust, and that it does not change how clinicians process specific cues. The "extremely familiar" cell is small ($n = 5$), so its mean should be read with caution.

A notable gender gap emerged in self-reported familiarity. Women reported significantly lower familiarity (median = 2, "slightly familiar") than men (median = 3, "moderately familiar"; Mann-Whitney $U = 7{,}164$, $p = .001$, rank-biserial $r = 0.22$, $n_{women} = 162$, $n_{men} = 113$). With codes 1 and 2 treated as "low familiarity," 58.0% of women fell in this band against 38.9% of men. The gap did not carry over into expressed trust. Once familiarity entered the model as a covariate, sex no longer predicted trust intercepts. The gap therefore plausibly reflects differences in prior exposure or technological self-efficacy, and it does not point to systematic differences in trust adjustment toward clinical AI. Professional experience showed a modest, non-significant relationship with trust in the multivariate model.

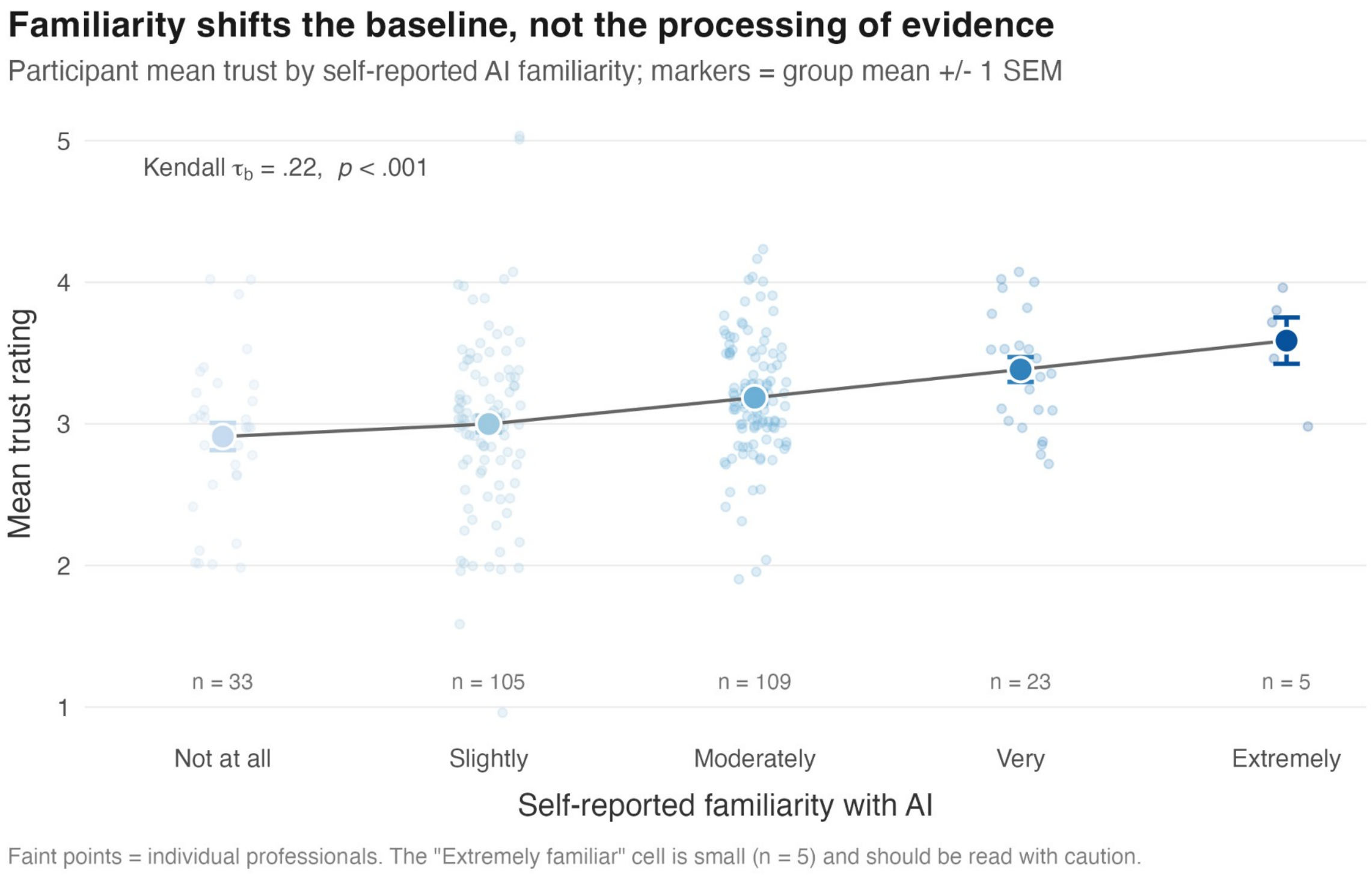

**Figure 4.** Participant-level mean trust as a function of self-reported AI familiarity. Faint points are individual professionals, and markers show the group mean ± 1 SEM. The monotonic increase indicates that familiarity raises baseline trust. The "extremely familiar" cell is small (n = 5).

### *3.4. Sensitivity analysis*

Analysis of vignette position revealed a slight fatigue effect ($\beta = -0.020$, $p = .002$) and no significant Position × Condition interactions. The experimental effects reported above are therefore not artefacts of presentation order (Table 2).

**Table 2.** Position-effect analyses.

| **Analysis** | **Statistic** | **Value** | ***p*** |
|---|---|---|---|
| Position–trust correlation | Kendall's τ | −0.073 | < .001 |
| Position coefficient (LMM) | β | −0.020 | .002 |

### *3.5. Predictive modelling*

Four algorithms trained on SPOT-engineered features were compared under participant-wise cross-validation (Table 3). LightGBM achieved the best performance ($R^2 = 0.06$, RMSE = 0.79), though all models performed similarly. The low $R^2$ under rigorous group-wise cross-validation marks a sharp drop from the fit obtained when individual baselines are available, and it shows that trust baselines are highly idiosyncratic. The LMM shows that the experimental factors reliably shift trust around a user's baseline. The baseline itself varies widely across individuals and is hard to predict from demographics alone.

Feature-importance analysis revealed two patterns. First, participant-level variables (professional role, experience, AI familiarity) were the strongest predictors of absolute trust scores. In other words, the prior P(Z), or who the clinician is, largely sets the starting point. Second, among the experimental manipulations, process signals (interface, consensus) consistently outweighed outcome signals, which mirrors the LMM results.

**Table 3.** Predictive-model performance under participant-level cross-validation.

| Model | $R^2$ (CV) | RMSE | MAE |
|---|---|---|---|
| LightGBM | 0.06 | 0.79 | 0.70 |
| XGBoost | 0.05 | 0.80 | 0.71 |
| Random Forest | 0.04 | 0.82 | 0.72 |
| Linear Regression | 0.03 | 0.83 | 0.73 |

### *3.6. Validation of the SPOT framework*

The critical test of SPOT concerns the relative weight of process and outcome signals. Drawing on rational-agent models, we had initially predicted outcome dominance of approximately 70% outcome and 30% process ($\beta > \alpha$). To estimate the empirical weights, we computed the share of total effect magnitude attributable to each class from the absolute LMM coefficients ($|\beta|$). We grouped interface, team consensus, explainability, autonomy, and personalization as process-related, and prognosis, accuracy, and intervention timing as outcome-related.

The results sharply contradicted the prediction and revealed a near-complete inversion of weights (Table 4; Figure 5). Healthcare professionals weighted process signals, namely transparency, social endorsement, and institutional legitimacy, roughly four times as heavily as outcome signals such as accuracy, expected benefit, and prognosis. This pattern suggests that clinicians depart from the purely consequentialist reasoning that rational-agent models predict. They appear to apply something like a theory of mind for machines, and they evaluate AI on dimensions that matter for intentional agents as well as on expected performance.

**Table 4.** Predicted and empirical signal weights under the SPOT framework.

| Signal type | Predicted (rational agent) | Empirical (LMM $\beta$) |
|---|---|---|
| Process signals | 30% | 80.3% |
| Outcome signals | 70% | 19.7% |

*Note. Empirical weights = sum of absolute LMM coefficients per category divided by the total.*

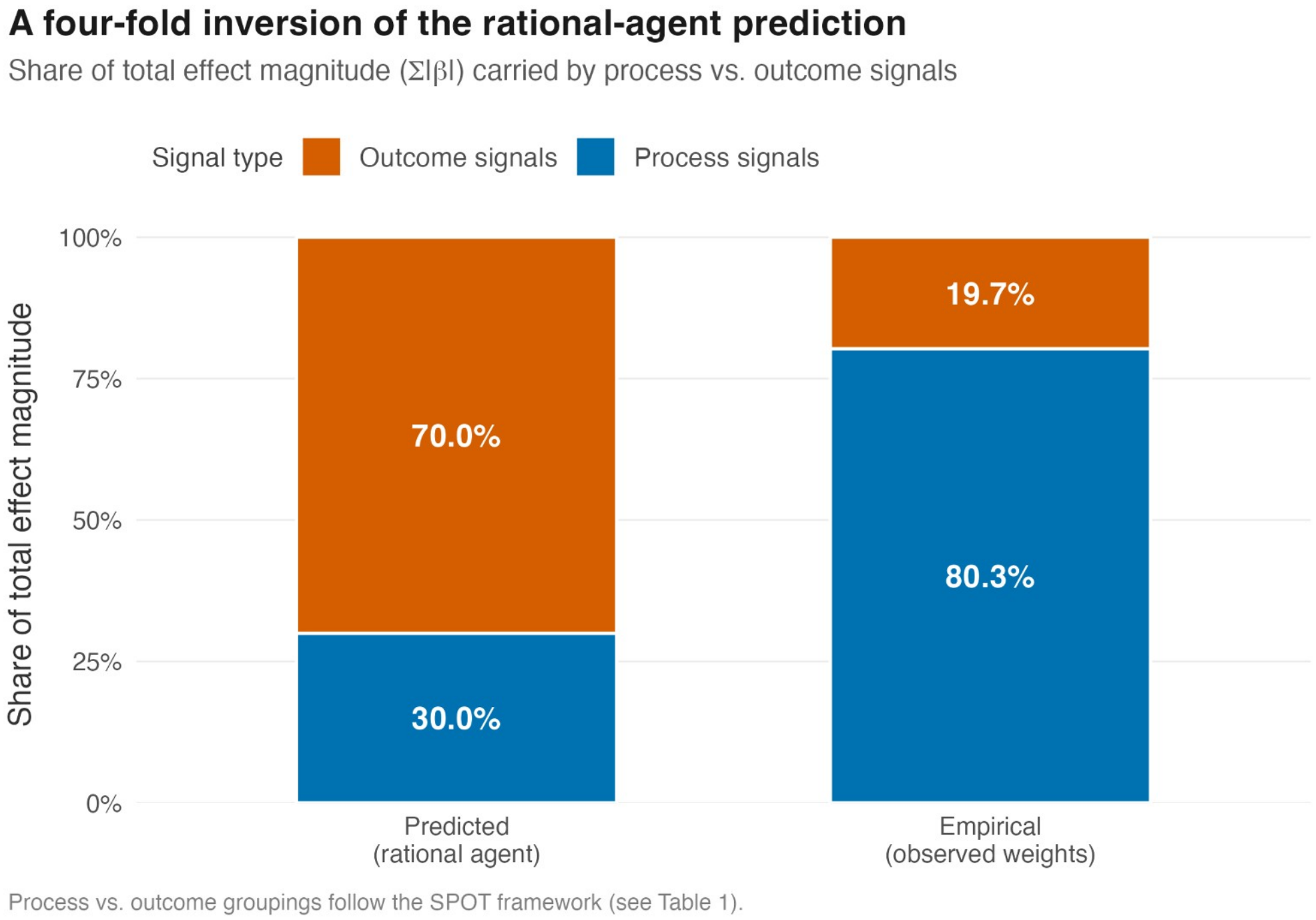

**Figure 5.** Predicted and empirical share of total effect magnitude (Σ|β|) carried by process and outcome signals. Rational-agent models predict outcome dominance near 70% outcome. The data show process signals carrying 80.3% of the weight, an approximately four-fold inversion.

## 4. Discussion

This study provides experimental evidence that healthcare professionals form trust judgments about AI through a process that looks more like social cognition than like dispassionate cost-benefit analysis. Process signals dominated, especially team consensus, interface familiarity, and explainability. This dominance suggests that clinicians adopt something like an intentional stance toward AI and treat these systems as agents whose trustworthiness needs social validation as well as technical verification.

The strong effect of team consensus ($\beta = 0.64$) is especially revealing. From a strictly rational standpoint, a colleague's opinion adds no information beyond the system's own performance metrics, because an AI that achieves 95% accuracy achieves it whether or not the team endorses it. Endorsement nonetheless produced a large increase in trust. We propose four non-exclusive mechanisms that fit the data. The first is epistemic outsourcing, where clinicians treat peer endorsement as evidence that people they trust have already vetted the system. The second is intention inference, where endorsement signals that the system aligns with professional values, so colleagues read an endorsed AI as one of their own. The third is risk distribution, where adopting a collectively endorsed system dilutes individual accountability for a bad outcome. The fourth is social learning, where social information drives rapid heuristic updating when direct outcome evidence is absent. Mechanisms one and four fit a deflationary, fully Bayesian account in which process cues act as diagnostic proxies for unobserved competence. Mechanisms two and three require the additional theoretical commitments of a genuine intentional-stance reading.

The explainability premium ($\beta = 0.63$) carries direct implications for design and regulation. Articles 13 (transparency) and 14 (human oversight) of Regulation (EU) 2024/1689 require that high-risk AI systems be understandable enough to enable meaningful human supervision, and our findings indicate that clinicians' trust adjustment already anticipates these requirements behaviourally. Participants treated epistemic opacity as a substantial penalty. They valued a system's capacity to give reasons as a near-prerequisite for partnership, and they treated black-box recommendations as a violation of the communicative norms expected of a clinical colleague, well beyond a mere technical limitation.

The robust effect of prognostic congruence ($\beta = 0.38$, favouring the adverse-prognosis vignette) indicates that clinicians evaluate AI consistency against their own clinical priors. Predictions that seem too good to be true undermine trust and do not enhance it. A system whose predictions clash with clinical expectation reads as miscalibrated or naive. The design corollary is clear. Systems may need to acknowledge typical population outcomes and quantify uncertainty before they offer optimistic predictions, so that optimism is not read as a signal of incompetence.

Our initial prediction of outcome dominance was disconfirmed. The empirical finding of process dominance ($\alpha \approx 0.80$) admits two readings. The first is deflationary. Clinicians remain approximately Bayesian and rationally assign more weight to process cues because those cues are more diagnostic of trustworthiness in this setting. If $P(\text{Process} \mid \text{Trustworthy})$ is more sharply peaked than $P(\text{Outcome} \mid \text{Trustworthy})$, so that process features discriminate trustworthy from untrustworthy systems more reliably than a nominal accuracy figure, then process signals should receive more weight in the posterior. This reading fits appropriate reliance under information scarcity (Lee & See, 2004) and computational-rationality accounts in which agents allocate limited inferential effort to the most informative cues (Gershman et al., 2015). The second reading is richer. Here clinicians apply a full theory of mind to the machine and evaluate its character, communicative conduct, and social standing, well beyond tracking competence alone. The two readings can both hold, and the present design cannot fully separate them. Separating them is an important target for future work, ideally with a hierarchical Bayesian model that estimates cue diagnosticities directly.

Several institutional implications follow. First, organizations should attend to social legitimacy as carefully as to technical performance, and they should introduce AI through team-based training and visible governance so that it enters practice as a collective, supported tool. Second, interface familiarity deserves strategic priority, and embedding AI within established workflows should precede the introduction of novel interaction paradigms. Third, AI systems should provide clinically meaningful and audience-appropriate explanations, and they should avoid post-hoc rationalizations. Fourth, performance claims should be calibrated to clinical priors to avoid triggering miscalibration signals. Fifth, disparities in AI familiarity, including the gender gap documented here, warrant targeted training. Familiarity shifts baseline trust and leaves the processing of evidence largely unchanged, so exposure may be enough to address it.

Several limitations qualify these conclusions. The vignette methodology captures cold-start trust formation, and it says little about how trust evolves through repeated interaction or recovers after an AI error. The Spanish, predominantly physician sample limits cross-cultural and inter-professional generalizability. Norms regarding professional authority and team hierarchy may differ elsewhere, and the under-representation of nursing (7.6%) is a notable gap, because nursing staff often interact most frequently with bedside AI. The low predictive $R^2$ (0.06) under participant-wise cross-validation underscores the idiosyncrasy of baseline trust and suggests that personalized trust modelling may be needed for effective deployment. The study was not pre-registered, so the analyses are best regarded as exploratory. The design, hypotheses, and analytic plan, together with anonymized data and scripts, have been deposited at OSF to support direct replication. Finally, we measured attitudinal trust, and behavioural reliance remains to be tested. Whether the observed explainability premium translates into safer clinical decisions is an open question.

**5. Concluding remarks**

Trust in clinical AI appears to function more like trust in a human agent than like reliance on a mechanical tool. Healthcare professionals form trust judgments by inferring latent properties of AI systems, such as their competence, intentions, and social standing, from observable cues that extend well beyond technical performance. The SPOT framework shows that process signals (transparency, consensus, familiarity) carry nearly four times the weight of outcome signals (accuracy, prognosis), which contradicts rational-agent models that predict outcome dominance.

These findings challenge assumptions embedded in both AI system design and regulatory frameworks. Technical accuracy is necessary, and on its own it is not sufficient for trust. Transparency helps, and it still requires social validation to cross the threshold needed for high-stakes decisions. Team consensus and institutional endorsement may matter as much as demonstrated performance. The findings also invite us to reconsider what it means for a clinician to trust an AI. Trust may rest above all on whether the system is the kind of agent that belongs in clinical practice, one that is transparent about its reasoning, endorsed by trusted peers, and embedded within legitimate governance. Building trustworthy clinical AI is therefore a social and institutional task as well as a technical one.

**Data availability**

The data underlying this article, together with the analysis scripts and the experimental materials, are available at the Open Science Framework (OSF). The reproducibility package includes the de-identified response data, the survey instrument, and supplementary materials. These data were collected within a broader research programme, and a companion study draws on the same deposit (see Section 2.4). Anyone requesting supplementary materials for this work should contact the authors.

## References

Aguinis, H., & Bradley, K. J. (2014). Best practice recommendations for designing and implementing experimental vignette methodology studies. *Organizational Research Methods*, *17*(4), 351–371. https://doi.org/10.1177/1094428114547952

ALLEA. (2023). *The European code of conduct for research integrity* (Rev. ed.). https://allea.org/code-of-conduct/

Asan, O., Bayrak, A. E., & Choudhury, A. (2020). Artificial intelligence and human trust in healthcare: Focus on clinicians. *Journal of Medical Internet Research*, *22*(6), e15154. https://doi.org/10.2196/15154

Baker, C. L., Saxe, R., & Tenenbaum, J. B. (2009). Action understanding as inverse planning. *Cognition*, *113*(3), 329–349. https://doi.org/10.1016/j.cognition.2009.07.005

CERN. (2020). *CERN open data policy for the LHC experiments*. https://doi.org/10.17181/CERN.QXNK.8L2G

Cheng, H. G., & Phillips, M. R. (2014). Secondary analysis of existing data: Opportunities and implementation. *Shanghai Archives of Psychiatry*, *26*(6), 371–375.

CSIC. (2021). *Código de buenas prácticas científicas del CSIC*. https://www.csic.es/sites/default/files/2023-01/cbpc_csic2021.pdf

Dennett, D. C. (1987). *The intentional stance*. MIT Press.

Dietvorst, B. J., Simmons, J. P., & Massey, C. (2015). Algorithm aversion: People erroneously avoid algorithms after seeing them err. *Journal of Experimental Psychology: General*, *144*(1), 114–126. https://doi.org/10.1037/xge0000033

Gershman, S. J., Horvitz, E. J., & Tenenbaum, J. B. (2015). Computational rationality: A converging paradigm for intelligence in brains, minds, and machines. *Science*, *349*(6245), 273–278. https://doi.org/10.1126/science.aac6076

Goddard, K., Roudsari, A., & Wyatt, J. C. (2012). Automation bias: A systematic review of frequency, effect mediators, and mitigators. *Journal of the American Medical Informatics Association*, *19*(1), 121–127. https://doi.org/10.1136/amiajnl-2011-000089

Greenhalgh, T., Wherton, J., Papoutsi, C., Lynch, J., Hughes, G., A'Court, C., Hinder, S., Fahy, N., Procter, R., & Shaw, S. (2017). Beyond adoption: A new framework for theorizing and evaluating nonadoption, abandonment, and challenges to the scale-up, spread, and sustainability of health and care technologies. *Journal of Medical Internet Research*, *19*(11), e367. https://doi.org/10.2196/jmir.8775

Griffiths, T. L., Chater, N., Kemp, C., Perfors, A., & Tenenbaum, J. B. (2010). Probabilistic models of cognition: Exploring representations and inductive biases. *Trends in Cognitive Sciences*, *14*(8), 357–364. https://doi.org/10.1016/j.tics.2010.05.004

Hidalgo, C. A., Orghian, D., Canals, J. A., de Almeida, F., & Martin, N. (2021). *How humans judge machines*. MIT Press. https://doi.org/10.7551/mitpress/13373.001.0001

Jara-Ettinger, J., Gweon, H., Schulz, L. E., & Tenenbaum, J. B. (2016). The naïve utility calculus: Computational principles underlying commonsense psychology. *Trends in Cognitive Sciences*, *20*(8), 589–604. https://doi.org/10.1016/j.tics.2016.05.011

Jian, J.-Y., Bisantz, A. M., & Drury, C. G. (2000). Foundations for an empirically determined scale of trust in automated systems. *International Journal of Cognitive Ergonomics*, *4*(1), 53–71. https://doi.org/10.1207/S15327566IJCE0401_04

Ke, G., Meng, Q., Finley, T., Wang, T., Chen, W., Ma, W., Ye, Q., & Liu, T.-Y. (2017). LightGBM: A highly efficient gradient boosting decision tree. *Advances in Neural Information Processing Systems*, *30*, 3146–3154.

Lee, J. D., & See, K. A. (2004). Trust in automation: Designing for appropriate reliance. *Human Factors*, *46*(1), 50–80. https://doi.org/10.1518/hfes.46.1.50_30392

Logg, J. M., Minson, J. A., & Moore, D. A. (2019). Algorithm appreciation: People prefer algorithmic to human judgment. *Organizational Behavior and Human Decision Processes*, *151*, 90–103. https://doi.org/10.1016/j.obhdp.2018.12.005

Longoni, C., Bonezzi, A., & Morewedge, C. K. (2019). Resistance to medical artificial intelligence. *Journal of Consumer Research*, *46*(4), 629–650. https://doi.org/10.1093/jcr/ucz013

Max Planck Society. (2021). *Rules of good scientific practice*. https://www.mpg.de/199493/regelnWissPraxis.pdf

Mayer, R. C., Davis, J. H., & Schoorman, F. D. (1995). An integrative model of organizational trust. *Academy of Management Review*, *20*(3), 709–734. https://doi.org/10.2307/258792

Rousseau, D. M., Sitkin, S. B., Burt, R. S., & Camerer, C. (1998). Not so different after all: A cross-discipline view of trust. *Academy of Management Review*, *23*(3), 393–404. https://doi.org/10.5465/amr.1998.926617

Seabold, S., & Perktold, J. (2010). Statsmodels: Econometric and statistical modeling with Python. In *Proceedings of the 9th Python in Science Conference* (pp. 92–96). https://doi.org/10.25080/Majora-92bf1922-011

Stoet, G. (2010). PsyToolkit: A software package for programming psychological experiments using Linux. *Behavior Research Methods*, *42*(4), 1096–1104. https://doi.org/10.3758/brm.42.4.1096

Stoet, G. (2017). PsyToolkit: A novel web-based method for running online questionnaires and reaction-time experiments. *Teaching of Psychology*, *44*(1), 24–31. https://doi.org/10.1177/0098628316677643

Tenenbaum, J. B., Kemp, C., Griffiths, T. L., & Goodman, N. D. (2011). How to grow a mind: Statistics, structure, and abstraction. *Science*, *331*(6022), 1279–1285. https://doi.org/10.1126/science.1192788

UNESCO. (2021). *Recommendation on open science*. https://unesdoc.unesco.org/ark:/48223/pf0000381148

Wilkinson, M. D., Dumontier, M., Aalbersberg, I. J., Appleton, G., Axton, M., Baak, A., Blomberg, N., Boiten, J.-W., da Silva Santos, L. B., Bourne, P. E., Bouwman, J., Brookes, A. J., Clark, T., Crosas, M., Dillo, I., Dumon, O., Edmunds, S., Evelo, C. T.,

Finkers, R., . . . Mons, B. (2016). The FAIR guiding principles for scientific data management and stewardship. *Scientific Data*, *3*, 160018. https://doi.org/10.1038/sdata.2016.18